\documentclass[sigconf,nonacm]{acmart}

\copyrightyear{2026}
\acmYear{2026}
\acmConference[SAO Workshop at CAIS '26]{AI Agents and Data Systems at the ACM Conference on AI and Agentic Systems (CAIS)}{May 26, 2026}{San Jose, California, USA}

\makeatletter
\AtBeginDocument{%
  \fancypagestyle{standardpagestyle}{%
    \fancyhf{}%
    \renewcommand{\headrulewidth}{\z@}%
    \renewcommand{\footrulewidth}{\z@}%

    \fancyfoot[C]{\if@ACM@printfolios\footnotesize\thepage\fi}%

    \fancyhead[LO]{\ACM@linecountL\@headfootfont\shorttitle}%
    \fancyhead[RE]{\@headfootfont\@shortauthors\ACM@linecountR}%

    \fancyhead[LE]{\ACM@linecountL\@headfootfont
      \acmConference@shortname, \acmConference@date, \acmConference@venue}%
    \fancyhead[RO]{\@headfootfont
      \acmConference@shortname, \acmConference@date, \acmConference@venue
      \ACM@linecountR}%
  }%
  \pagestyle{standardpagestyle}%
}
\makeatother

\usepackage{multirow}
\usepackage{enumitem}
\usepackage{arydshln}
\usepackage{makecell}
\usepackage[capitalize,noabbrev]{cleveref}
\begin{document}

\newcommand\blfootnote[1]{%
  \begingroup
  \renewcommand\thefootnote{}%
  \footnotetext{#1}%
  \addtocounter{footnote}{-1}%
  \endgroup
}

\title{Beyond Semantic Similarity: Performance and Costs of Agentic Retrieval for Complex Tasks}

\author{%
  Reza Esfandiarpoor\textsuperscript{*\,1},
  Radek Osmulski\textsuperscript{*\,1},
  Yauhen Babakhin\textsuperscript{1},
  Gabriel de Souza P. Moreira\textsuperscript{1},
  Oliver Holworthy\textsuperscript{1},
  Jie He\textsuperscript{\ddag\,2},
  Ronay Ak\textsuperscript{1},
  Jiarui Cai\textsuperscript{1},
  Ryan Chesler\textsuperscript{1},
  Bo Liu\textsuperscript{1},
  Even Oldridge\textsuperscript{1}%
}
\affiliation{%
  \institution{\textsuperscript{1}NVIDIA \quad \textsuperscript{2}University of Edinburgh}
  \country{}
}

\renewcommand{\shortauthors}{Esfandiarpoor et al.}

\begin{abstract}
Modern information systems, including many agentic workflows, use dense retrieval to explore large amounts of unstructured data.
However, dense retrieval relies on surface-level semantic similarity, which is insufficient for increasingly complex search applications.
Here, we investigate agentic retrieval that combines the reasoning capabilities of Large Language Models (LLMs) with the efficient corpus exploration of retrievers in a ReAct agentic loop to solve complex retrieval tasks.
In our experiments, we show that agentic retrieval is more effective than standard retrieval, improving nDCG@10 by 8.7 points using the same embedding model.
Moreover, while specialized retrieval methods struggle on out-of-domain tasks, agentic retrieval is highly generalizable: the same pipeline achieves competitive results on both the ViDoRe v3 and BRIGHT leaderboards.
However, this improvement comes at a cost.
On average, agentic retrieval takes 107.4 seconds, compared to 0.67 seconds for standard retrieval, and consumes 764.1K input and 5.8K output tokens per query.
In short, our study demonstrates the effectiveness of agentic retrieval in modern data systems and motivates future work on more cost-efficient retrieval agents for large-scale deployment.
Code: \url{https://github.com/NVIDIA/NeMo-Retriever/tree/main/retrieval-bench}

\end{abstract}




\maketitle

\renewcommand{\thefootnote}{\fnsymbol{footnote}}
\footnotetext[1]{Equal contribution.}            
\footnotetext[3]{Work done during internship at NVIDIA.}   
\renewcommand{\thefootnote}{\arabic{footnote}}

\setcounter{footnote}{0}

\section{Introduction}
\label{sec:intro}

Many AI workflows, such as retrieval-augmented generation~\citep{lewis2020rag,shi2023replug} and DeepResearch~\citep{chen2025browsecomp}, use information retrieval to process massive amounts of unstructured data.
With the growing popularity of these applications, retrieval tasks are shifting from targeted, well-defined search queries to high-level and abstract task descriptions~\citep{chen2025browsecomp,esfandiarpoor2025themcpcompany}.
For example, an agent given a math problem should be able to retrieve potentially useful theorems based on the problem description~\citep{su2024bright}.
Here, we study agentic retrieval to address the needs of complex and reasoning-intensive retrieval tasks\footnote{An earlier version of this work has been previously released as the NVIDIA NeMo Retriever Agentic Retrieval pipeline~\citep{osmulski2026agenticretrieval}.}.

\blfootnote{\textit{SAO Workshop at CAIS '26, May 26, 2026, San Jose, California, USA.}}
\setcounter{footnote}{1}

The needs of these emerging retrieval tasks go beyond the capabilities of standard dense retrieval, which relies solely on semantic similarity represented by the cosine similarity between vector representations of the query and documents~\citep{karpukhin2020dense}.
In more complex retrieval tasks, however, there may be limited surface-level semantic similarity between a given query, such as a math problem, and the relevant documents, such as useful mathematical theorems.
Instead, successfully completing these tasks requires higher-level skills, including reasoning, such as searching over mathematical theorems; knowledge of the dynamics of real-world systems, such as retrieval of tool specifications; and iterative exploration, such as in DeepResearch~\citep{su2024bright,esfandiarpoor2025themcpcompany,chen2025browsecomp}.

In this work, we investigate an agentic approach to information retrieval that combines the capabilities of LLMs and dense retrievers through a ReAct loop~\citep{yao2023react} to address the requirements of complex retrieval tasks.
Our agent inherits the reasoning skills and world knowledge of LLMs while also being able to sift through a massive corpus using lightweight dense retrievers.
We evaluate our agent on two dataset collections: ViDoRe v3~\citep{loison2026vidorev3comprehensiveevaluation} for enterprise document retrieval and BRIGHT~\citep{su2024bright} for reasoning-intensive retrieval tasks.
Our main findings are:

\begin{itemize}
    \item \textbf{Effectiveness of Agentic Retrieval}.
    We show that agentic retrieval is a promising approach for solving complex retrieval tasks.
    On average, our best agent achieves 8.7 points better nDCG@10 than standard retrieval with the same embedding model.
    \item \textbf{Generalizability}.
    Agentic retrieval generalizes effectively across domains without any changes.
    While specialized methods experience a significant drop in performance on out-of-domain tasks, our agentic retrieval pipeline achieves competitive results on both the ViDoRe v3 and BRIGHT leaderboards without modification.
    \item \textbf{Costs}.
    In addition to performance, we study the costs associated with agentic retrieval, in terms of additional compute.
    We find that agentic retrieval incurs significantly higher costs than standard dense retrieval, and further work is needed before retrieval agents can be deployed at scale.
\end{itemize}

Our work provides empirical evidence that standard dense retrieval is insufficient for emerging retrieval applications and establishes agentic retrieval as a promising direction.
We also discuss the limitations of widespread adoption of agentic retrieval, primarily in terms of cost and efficiency.
We hope our results encourage future work on agentic retrieval that addresses these limitations.

\begin{figure*}[t]
  \centering
  \includegraphics[height=5.5cm, trim=0 60 0 40, clip]{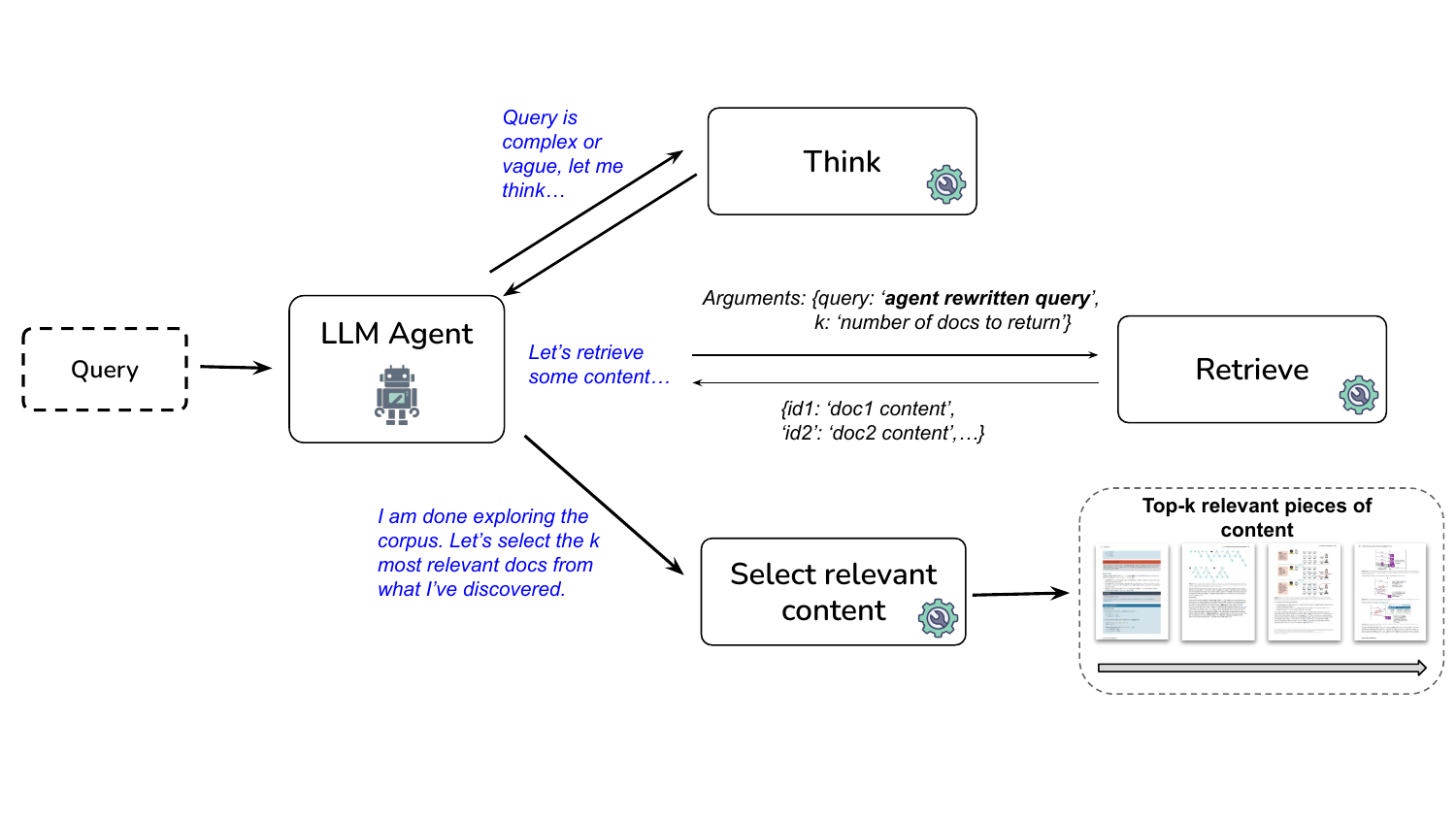}
  \caption{Overview of our retrieval agent. Given a query,
  the agent iteratively invokes the \texttt{think} tool to reason and plan for complex queries,
  the \texttt{retrieve} tool to explore the corpus,
  and finally \texttt{final\_results} to return the top-$k$ discovered documents that are most relevant to the query.}
  \label{fig:pipeline}
\end{figure*}
\section{Retrieval Agent}
\label{sec:method}

The skills needed to solve information-seeking tasks effectively are spread across two types of models.
On one hand, LLMs possess significant reasoning capabilities and world knowledge, but they struggle to handle massive corpora of millions of documents.
On the other hand, retrievers can easily sift through millions of documents, but they are often small and lack extensive reasoning capabilities.
Prior attempts to combine these two model types often rely on static, manually designed workflows.
For example, in query rewriting, an LLM is used to rewrite the original query into one or more revised queries, which are then used as inputs to the retriever~\citep{wang2023query2doc,jagerman2023query}.
This static design limits the autonomy and capabilities of the final solution.
For instance, in such designs, the LLM cannot refine queries based on newly discovered information from the retriever.

To address this issue, we create NeMo Retriever Agent in which the query is given to the LLM, and the LLM can then invoke the retriever in a loop to explore the corpus as needed until it discovers the necessary information (\cref{fig:pipeline}).
This design allows the LLM to adapt its search strategy to each query and corpus, and to dynamically decide when, how often, and with what queries the retriever should be called.
We implement this agent using a ReAct loop and standard tool calling through JSON schemas~\citep{yao2023react}.
Compared with custom prompting templates, the ReAct loop makes the agent easier to extend in the future by adding new tools, such as tools for using different retrievers simultaneously, and also enables feedback to be provided to the agent through tool outputs.

Specifically, our agent starts with a system prompt that explains the goal of the task, namely discovering all relevant documents, and describes each tool.
The main query is provided in the first user message.
We also include the initial documents retrieved for the original query by the dense retriever as part of the user message, and ask the agent to find any missing documents.
These initial retrieval results avoid unnecessary search attempts and, more importantly, allow the agent to understand the type of documents in the corpus, such as math theorems or function descriptions, and adjust its search strategy and subsequent queries accordingly.
Finally, the agent is provided with the following tools that it can use to solve the task and report the results.

\begin{itemize}
    \item \textbf{Retrieve}.
    The Retrieve tool takes a query and an integer k as arguments.
    It is powered by a dense retriever and uses the cosine similarity between query and document embeddings to select the top-k documents most similar to the query and returns them to the agent, along with their unique IDs.
    
    \item \textbf{Think}.
    The Think tool takes a string thought argument, which allows the agent to generate additional tokens for reasoning about complex queries or planning its search strategy.
    The Think tool does not return any output or change the environment.

    \item \textbf{Final\_Results}.
    This tool allows the agent to report the results and terminate the interaction.
    It expects as input a list of k document IDs that are most similar to the query, sorted by relevance.
    If the agent selects the wrong number of documents, the tool returns an error and asks the agent to try again.
    This tool also expects a second argument containing the agent's rationale for selecting these documents.
\end{itemize}

\paragraph{Reliability}
Unlike dense retrievers, retrieval agents can fail for a variety of reasons before selecting the final list of documents.
Such failures include exceeding the context window or raising content violation errors.
To improve reliability, in these cases, our pipeline falls back to Reciprocal Rank Fusion (RRF).
Specifically, when an error occurs, we collect the ranked list of documents from each search attempt before the error.
We then use RRF to merge these rankings and calculate a final score for each of these documents.

\paragraph{Infrastructure}
Model Context Protocol (MCP) has emerged as a standard method for exposing tools to LLMs~\citep{anthropic2024mcp}.
However, in our experiments, we find that for retrieval agents, where a limited number of tools is called many times, MCP leads to compounding overheads: each run requires spinning up a separate server, loading corpus embeddings into GPU memory, and managing the lifecycle of both client and server processes.
Moreover, network round-trips add latency to every retrieval call.
To address these issues, we replace the MCP server with a \emph{thread-safe singleton retriever} that lives in the same process as the agent.
In this approach, the model and corpus embeddings are loaded once, all access is protected by a reentrant lock, and the same \texttt{retrieve()} interface is exposed to several concurrent agents.
This implementation eliminates an entire class of deployment errors and substantially improves GPU utilization and experiment throughput.

\section{Experiments}
\label{sec:experiments}

\subsection{Setup}

\paragraph{Datasets}
We evaluate our pipeline on two benchmarks for complex retrieval.
First, we use ViDoRe v3~\citep{loison2026vidorev3comprehensiveevaluation}, a benchmark for enterprise document retrieval that spans six languages and 10 domains, including finance, pharmaceutical, and government energy reports.
It also contains seven types of queries, such as extractive queries, multi-hop queries, and queries requiring numerical reasoning.
We also evaluate on the BRIGHT benchmark for reasoning-intensive text retrieval~\citep{su2024bright}.
Across different domains, BRIGHT measures complex reasoning skills, including logical deduction, code understanding, and mathematical reasoning.

\paragraph{Models}
We evaluate our agent using Opus 4.5 as a capable proprietary model.
We also experiment with gpt-oss-120b\footnote{\url{https://huggingface.co/openai/gpt-oss-120b}} as a representative open-source model.
For ViDoRe v3, we use the colembed-vl-8b-v2 retriever\footnote{\url{huggingface.co/nvidia/nemotron-colembed-vl-8b-v2}}~\citep{moreira2026nemotroncolembedv2topperforming}, a capable model for visual document retrieval.
For BRIGHT, we use llama-nv-reasoning-3b\footnote{\url{huggingface.co/nvidia/llama-nv-embed-reasoning-3b}}, which is trained specifically for reasoning tasks.
We also report results with the smaller llama-nemotron-1b\footnote{\url{huggingface.co/nvidia/llama-nemotron-embed-vl-1b-v2}} model on both datasets.

\begin{table}[t]
\centering
\caption{Performance of agentic retrieval and standard dense retrieval with different models.}
\label{tab:ablation}
\small
\begin{tabular}{lllc}
\toprule
\textbf{Method} & \textbf{LLM} & \textbf{Embedding} & \textbf{NDCG@10} \\
\midrule
\multicolumn{4}{c}{\textit{ViDoRe~v3}} \\
\midrule
\multirow{3}{*}{\shortstack[c]{NeMo Retriever\\Agent}}
& Opus 4.5 & colembed-vl-8b-v2 & \textbf{69.22} \\
& gpt-oss & colembed-vl-8b-v2 & 66.38 \\
& gpt-oss & llama-nemotron-1b & 62.42 \\

\multirow{2}{*}{\shortstack[c]{Standard Retrieval}}
 & - & colembed-vl-8b-v2 & 64.36 \\
 & - & llama-nemotron-1b & 55.83 \\
\midrule
\multicolumn{3}{c}{\textit{BRIGHT}} \\
\midrule
\multirow{3}{*}{\shortstack[c]{NeMo Retriever\\Agent}}
 & Opus 4.5 & llama-nv-reasoning-3b & \textbf{50.79} \\
 & gpt-oss & llama-nv-reasoning-3b & 41.27 \\
 & gpt-oss & llama-nemotron-1b & 33.85 \\
\multirow{2}{*}{\shortstack[c]{Standard Retrieval}}
& - & llama-nv-reasoning-3b & 38.28 \\
& - & llama-nemotron-1b & 19.56 \\
\bottomrule
\end{tabular}
\end{table}

\subsection{Results}
\Cref{tab:ablation} reports the performance of agentic and standard retrieval with different models.
Across all cases, agentic retrieval outperforms the same embedding model used in a standard retrieval pipeline.
This shows that our retrieval agent better understands the complexity of the given queries and uses the embedding models more effectively than standard dense retrieval pipelines.
Our results also emphasize the importance of reasoning capabilities in modern retrieval problems: using the best embedding model for each dataset, our agent with the stronger Opus 4.5 model achieves, on average, a 6.2 point higher nDCG@10 than the same pipeline with gpt-oss-120b.
This performance difference is more pronounced on the BRIGHT benchmark, which requires more extensive reasoning skills, where the gap is 9.5 nDCG@10 points.
In our investigation on ViDoRe v3, we find that Opus 4.5 makes more search calls per query, 9.2 on average, than gpt-oss-120b, which makes 2.4 calls on average.
This suggests that the limited exploration capabilities of gpt-oss-120b may be one cause of the performance difference.

A particularly interesting finding is that the agent can partially compensate for the limitations of the embedding model.
Specifically, the performance gap between weaker and stronger embedding models is much smaller in the agentic retrieval setup than in standard retrieval with the same models: agentic retrieval reduces the performance gap between embedding models with different capabilities by almost half, 57\%, on average.
This is particularly important because embedding models are often kept very small to handle large numbers of documents, which limits their capabilities.

\subsection{Analysis}
\label{sec:analysis}

\paragraph{Generalizability}
Agentic retrieval, with its dynamic search strategy for each task, generalizes better to a wide range of tasks than specialized retrieval systems.
As of March 13, 2026, our retrieval agent holds the \#1 spot on the ViDoRe v3 leaderboard\footnote{\url{huggingface.co/spaces/vidore/vidore-leaderboard?tab=vidore-v3-pipeline}} and the \#2 spot on the BRIGHT leaderboard\footnote{\url{github.com/NVIDIA/NeMo-Retriever/blob/main/retrieval-bench/submissions/bright_agentic.md}}.
To put the generalizability of agentic retrieval in context, we evaluate the \#1 method from the BRIGHT leaderboard, INF-X-Retriever~\citep{infxretriever2025}, on the ViDoRe v3 benchmark and find that its performance is significantly lower than that of our retrieval agent (\cref{tab:crossbenchmark}).
Moreover, we also evaluate their pipeline using the same retriever that our agent uses.
Although the new embedding model improves its performance, the specialized method for the BRIGHT benchmark performs even worse than the standard retrieval baseline in this case.
In contrast, the same retrieval agent, without any changes, achieves competitive results on both benchmarks, demonstrating that the agent's dynamic adaptation provides genuine cross-domain generalizability.

\begin{table}[t]
\centering
\caption{Performance of our agent and INF-X-Retriever~\citep{infxretriever2025} on ViDoRe v3 and BRIGHT. We also report the performance of standard dense retrieval as well as that of INF-X-Retriever using the same embedding model as our agent. \#1 and \#2 are leaderboard ranks on March 13, 2026.}
\label{tab:crossbenchmark}
\small
\begin{tabular}{lcc}
\toprule
\textbf{Pipeline} & \textbf{ViDoRe v3} & \textbf{BRIGHT} \\
\midrule
NeMo Retriever Agent (ours) & \textbf{69.22} (\#1) & 50.90 (\#2) \\
INF-X-Retriever & 51.01 & \textbf{63.40} (\#1) \\
\midrule
INF-X + nemotron-colembed-vl-8b-v2 & 62.31 & -- \\
Dense only (nemotron-colembed-vl-8b-v2) & 64.36 & -- \\
\bottomrule
\end{tabular}
\end{table}

\paragraph{Successful Search Patterns}
The combination of the LLM's reasoning skills and the iterative agentic loop leads to several recurring search patterns that contribute to the agent's success.
\emph{Query Refinement}: the agent dynamically refines and adjusts its subsequent search queries based on newly discovered information.
\emph{Persistent Rephrasing}: the agent persistently rephrases queries until it finds useful information.
\emph{Complexity Decomposition}: the agent breaks complex, multi-part queries into multiple simpler queries with clear goals, which are easier for the embedding model to understand.

\begin{table}[t]
\centering
\caption{Average delay and costs of agentic retrieval on ViDoRe v3 for each query.}
\label{tab:cost}
\small
\begin{tabular}{lcc}
\toprule
& \textbf{Opus 4.5} & \textbf{gpt-oss-120b} \\
\midrule
Delay (sec.) & 136.3 & 78.6 \\
Input tokens & 759.4k & 768.9k \\
Output tokens & 6.2k & 5.4k \\
\bottomrule
\end{tabular}
\vspace{-3mm}
\end{table}

\paragraph{Costs}
The performance gains of agentic retrieval come at a cost.
\Cref{tab:cost} reports the average costs of agentic retrieval for each query in ViDoRe v3.
While standard retrieval takes 0.67 seconds, agentic retrieval takes 107.45 seconds per query on average.
Moreover, completing each query consumes 764.1k input and 5.8k output tokens, which is an additional cost compared to standard retrieval.
The additional overhead of LLM inference is the main limitation of agentic retrieval at scale.
However, given the increasing complexity of retrieval tasks and the ability of retrieval agents to address this complexity, we hope our results encourage future work on more efficient models and approaches for agentic retrieval.

\section{Discussion}
\label{sec:discussion}

Our work offers several lessons for future work on data systems that handle massive amounts of unstructured data.

\paragraph{Evolving applications of retrieval demand new approaches to retrieval}
While standard retrieval based on semantic similarity works well for narrow and targeted search queries, it is insufficient for emerging retrieval applications in systems, such as RAG or DeepResearch.
These applications demand high-level skills, such as iterative exploration and intermediate reasoning steps, that go beyond what standard retrieval methods offer.
Our work shows that agentic retrieval is a promising approach for these more complex retrieval tasks, and our results encourage future work in this direction.

\paragraph{The costs of retrieval go beyond embedding and index lookup.}
Agentic retrieval introduces new dimensions to retrieval cost, namely the number of reasoning steps and tokens consumed, which are functions of query complexity and dynamic agent behavior.
This differs from standard retrieval, where costs primarily correlate with the size of the search index and, therefore, the size of the corpus.
Thus, just as there are different methods for controlling costs associated with the search index, such as variants of approximate nearest neighbor search~\citep{douze2024faiss}, future work on agentic retrieval should investigate similar capabilities for agentic retrieval, allowing users to adapt token and reasoning costs to their specific applications.

\paragraph{More efficient and effective open-source models for agentic retrieval}
As shown in \cref{tab:ablation}, the performance of open-source models noticeably lags behind that of proprietary counterparts.
Moreover, unlike general-purpose agents, the task for retrieval agents is clear, well defined, and requires very specific skills, such as iterative exploration.
Taken together, we believe a promising direction for future work is to develop novel models for agentic retrieval that are more effective than current open-source models and more efficient than proprietary models like Opus 4.5.

\paragraph{Infrastructure and system design should adapt to the needs of the target task}
As evidenced by our migration from an MCP-based retriever tool to an in-process implementation, standard approaches for agent development are not the best choice for all applications.
Another example is tool-calling patterns.
Agents often make sequential tool calls one at a time, which aligns with the nature of most real-world tasks.
For retrieval, however, parallel tool calls, that is, multiple search attempts in a single turn, are a better option and can reduce the number of steps while potentially improving results by providing more context for the next search attempt.
We believe such task-specific optimizations are key to creating effective and efficient retrieval agents which are reusable across many systems.

\subsection{Related Work}

Retrieval methods have long been based on semantic similarity between queries and documents as captured by their vector embeddings~\citep{qu2021rocketqa,karpukhin2020dense,xiong2020approximate,zhan2021optimizing}.
Many works have also used LLMs to improve this paradigm, for example by rewriting and improving the original query~\citep{jagerman2023query,wang2023query2doc,dhole2024genqrensemble}.
However, this line of work still relies on a static pipeline and does not fully take advantage of emerging LLM capabilities, such as iterative reasoning and exploration.
Agentic retrieval, in contrast, goes beyond semantic similarity alone and incorporates the agentic capabilities of recent LLMs in a dynamic pipeline that adapts to the complexity of each query.

In another line of work, retrievers are frequently incorporated as key components of modern agentic systems for solving specific tasks, such as DeepResearch~\citep{chen2025browsecomp,cheng2025agent,jin2025search}.
In these systems, the retrieval method remains the same as before, while the goal is to have an agent that can effectively use standard retrieval to solve a specific task.
Conversely, our goal is to make the retrieval task itself agentic and to create a generalizable retrieval agent that can be reused as the search component of many downstream applications.

\section{Conclusion}
\label{sec:conclusion}

In this work, we investigate agentic retrieval as a promising solution for addressing the increasing complexity of retrieval tasks.
Since standard dense retrieval is based solely on semantic similarity between query and document embeddings, we develop a ReAct agent that uses standard retrievers to explore the corpus while also using LLMs to provide the higher-level capabilities needed for complex queries, such as reasoning and iterative exploration.
Our experiments on two complex benchmarks, ViDoRe v3 and BRIGHT, show that agentic retrieval is better equipped to address this level of complexity and significantly improves performance compared to standard dense retrieval.
Moreover, we show that agentic retrieval is highly generalizable: the same pipeline performs well on diverse tasks, whereas the performance of specialized solutions drops significantly when they are evaluated on out-of-domain tasks.
Finally, we compare the costs of agentic and standard retrieval and discuss the additional compute requirements of retrieval agents.
Overall, our work demonstrates the promise of agentic retrieval, discusses its limitations, and encourages future work to further improve the efficiency and effectiveness of agentic retrieval.

\bibliographystyle{ACM-Reference-Format}
\bibliography{paper}

\end{document}